\documentclass[a4paper, 12 pt, conference, onecolumn]{ieeetran}  
\IEEEoverridecommandlockouts
\let\labelindent\relax

\usepackage[left=3cm,right=3cm,top=2.5cm,bottom=3cm]{geometry}

\usepackage{cite}
\usepackage{amsmath,amssymb,amsfonts,amsthm} 
\usepackage{mathtools}
\usepackage{graphicx}
\usepackage{eurosym}
\usepackage{textcomp}
\usepackage{xcolor}

\usepackage[linesnumbered, ruled]{algorithm2e}
\usepackage{url}
\usepackage{enumitem}
\usepackage{dsfont}
\usepackage[utf8x]{inputenc}
\usepackage{lmodern,textcomp}

\newtheorem{theorem}{Theorem}
\newtheorem{lemma}{Lemma}

\newtheorem{proposition}{Proposition}
\newtheorem{corollary}{Corollary}
\newtheorem{remark}{Remark}
\newtheorem{problem}{Problem}

\newcommand{\users}{\ensuremath{\mathcal{U}}}
\newcommand{\eps}{\ensuremath{\varepsilon}}
\newcommand{\maxenergy}{\ensuremath{\overline{e}}}
\newcommand{\maxEnergy}{\ensuremath{\overline{E}}}

\def\cT{\mathcal{T}}
\def\cS{\mathcal{S}}
\def\esc{E^{c^*}}
\def\esd{E^{d^*}} 
\def\etc{\widetilde E^c}
\def\etd{\widetilde E^d} 
\def\Lm{\mathcal{L}^-} 
\def\Lp{\mathcal{L}^+} 
\def\cT{\mathcal{T}} 

\title{On Optimal Management of Energy Storage Systems in Renewable Energy Communities}

\author{Giovanni Gino Zanvettor, Marco Casini, Antonio Vicino
	\thanks{This work was supported by: (i) Italian Ministry for Research in the framework of the 2022 Program for Research Projects of National Interest (PRIN) under Grant 2022K4CLL3; (ii) University of Siena in the framework of the Piano per lo Sviluppo per la Ricerca (PSR) 2023, F-NEW FRONTIERS.}%
	\thanks{The authors are with the Dept. of Information Engineering and Mathematics, University of Siena, Italy.\newline
		Email: \{\texttt{zanvettor,casini,vicino}\}@diism.unisi.it}}%
\begin{document}
	
	\maketitle
		\begin{abstract}
		Renewable energy communities are legal entities involving the association of citizens, organizations and local businesses aimed at contributing to the green energy transition and providing social, environmental and economic benefits to their members. This goal is pursued through the cooperative efforts of the community actors and by increasing the local energy self-consumption. In this paper, the optimal energy community operation in the presence of energy storage units is addressed. By exploiting the flexibility provided by the storage facilities, the main task is to minimize the community energy bill by taking advantage of incentives related to local self-consumption. Optimality conditions are derived, and an explicit optimal solution is devised. Numerical simulations are provided to assess the performance of the proposed solution.
	\end{abstract}

	\section{Introduction}
	The Net Zero plan of the European community aims at achieving climate neutrality by 2050 \cite{greenDeal}. One of the most promising solutions relies on the paradigm of renewable energy communities (RECs). As stated by the European Union \cite{EUDir944}, an energy community is a legal entity whose participation is established on a voluntary and open basis, with the primary purpose of ensure environmental, social and economic benefits to its members and shareholders, by providing auxiliary services through renewable generation facilities, energy storages and electric vehicles. In fact, it may contribute to reducing gas emissions, triggering renewable self-consumption mechanisms, and increasing the overall sustainability of the environment. 
	
	In this paper, we focus on incentive-based RECs, like in the Italian regulation, where incentive programs are promoting the self-consumption mechanism \cite{stentati2022optimization, cielo2021}. More specifically, the incentive is proportional to the virtual self-consumption of the community in a given time period, defined as the minimum between the overall energy demand of the community and the total renewable energy generated in the community. 
	Several studies are presented in the literature aimed at assessing the benefits provided by RECs \cite{Hoicka2021, Ceglia2021}, increasing community efficiency \cite{Martirano2021, Cutore2023}, and proposing incentive redistribution schemes \cite{stentati2023redistribution}. In \cite{battaglia2024}, the economic benefits of a REC are evaluated in the context of high efficiency buildings, whereas the performance of an energy community in the presence of a fleet of electric vehicles has been reported in \cite{zanvettor2022}. A game-theoretic approach to design an incentive mechanism and foster community self-consumption has been adopted in \cite{Lilliu2023}, economic analyses about the profitability of Italian energy communities according to different factors have been proposed in \cite{dadamo2024}, and optimization models for the optimal allocation of renewable resources have been developed in \cite{Sousa2023}.

	\subsection*{Contribution}
	In this paper, an incentive-based REC composed of consumers, prosumers and producers is considered. The incentive provided to the community is computed on the basis of the virtual self-consumption in each time period of the day. We suppose that prosumers and producers are equipped with energy storage units to provide flexibility inside the community. A central entity, called REC manager, is engaged to manage the community. Its main role is to coordinate the storage units of the community to minimize the community cost according to the profiles of load and renewable generation provided by each member. Thus, the optimal control problem is formulated and optimality conditions are derived to find the optimal storage schedule. 
	This paper contributes to the existing literature by devising an optimization framework to manage energy storage units inside an energy community. Most importantly, it provides analytical tools that streamline the resolution of the introduced problem, allowing for simplified analyses about the convenience and the potential benefits of energy storage systems in RECs.

	\subsection*{Paper organization}
	The remainder of this paper is structured as follows. In Section~\ref{sec:formulation}, the community model is presented, whereas the optimization problem and technical results are provided in Section~\ref{sec:optimization}. Illustrative examples and related results are reported in Section~\ref{sec:results}, while conclusions are drawn in Section~\ref{sec:conclusion}.
	
	\subsection*{Notation}
	For a given optimization problem, we denote with the superscript $^*$ all the quantities related to the optimal solution $\cS^*$. Accordingly, all the variables related to a given solution $\widetilde\cS$ are denoted with the superscript~ $\widetilde{}$~.

	\section{Problem Formulation}\label{sec:formulation}
	
	We suppose to work in a discrete time setting where the sampling time is denoted by $\Delta$.
	Let us focus on a renewable energy community composed of a set $\users$ of members (or entities). For a given community member $u\in\users$, let $l_u(t)$ be the load between time $t$ and $t+1$, whereas $r_u(t)$ be the renewable energy generated in the same time interval. 
	In general, an entity can be a consumer, a producer or a prosumer. A consumer is characterized by its load ($l_u(t)\ge0,~r_u(t)=0,~\forall t$), a producer by the generated renewable energy ($l_u(t)=0,~r_u(t)\ge0,~\forall t$), while a prosumer involves both consumption and generation ($l_u(t)\ge0,~r_u(t)\ge0,~\forall t$).
	
\begin{figure}[t!]
		\centering
		\includegraphics[width=0.6\columnwidth]{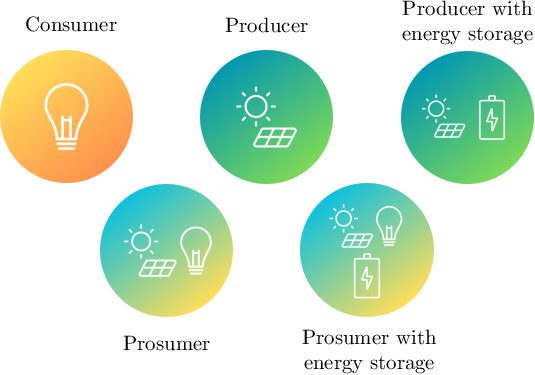}
		\caption{Entity types in the considered community.}\label{fig:Community_entities}
	\end{figure}

	Let us assume that a fraction of producers and prosumers are equipped with an energy storage, and let them be gathered into the set $\users_s\subseteq\users$. Consumers are not considered in this set because, as stated by the Italian energy service manager \cite{GSE-StorageRegulation}, a storage can contribute to the community only by using renewable sources. Thus, we will consider 5 kinds of entities, which represent all the possible combinations of load, generation and storage, as depicted in Fig.~\ref{fig:Community_entities}.
	
	Let $\rho_u(t) = r_u(t)-l_u(t)$; for a given entity $u\in\users_s$ the maximum amount of energy that can be charged into the storage between $t$ and $t+1$ is 
	\begin{equation*}
		\maxenergy^{c}_u(t) = \max\{\rho_u(t) ,0\}.
	\end{equation*}
	Note that $\maxenergy^c_u(t)$ denotes the surplus between the generated and the consumed energy; the remaining generated energy is supposed to supply the entity load. For simplicity, all the storage systems are assumed to have unlimited capacity and without technical limitations about maximum charging and discharging power rates.
	
	Given an entity $u\in\users_s$, let $e^c_u(t)$ be the energy charged into the storage while $e^d_u(t)$ be the energy drawn from the storage between $t$ and $t+1$. The energy stored at time $t+1$ is expressed as
	\begin{equation*}
		s_u(t+1) = s_u(t)+\eta e^c_u(t) - \frac{1}{\eta}e^d_u(t),
	\end{equation*}
	where $\eta<1$ denotes the battery efficiency.
	Clearly, $e^c_u(t)$ is a positive quantity and it is bounded by the surplus of generated energy, that is
	\begin{equation*}
		0\leq e^c_u(t)\leq\maxenergy^c_u(t)\quad\forall t.
	\end{equation*}
	On the other hand, since $s_u(t)\ge0$, $e^d_u(t)$ is positive and cannot exceed the stored energy 
	\begin{equation*}
		0\leq e^d_u(t)\leq \eta s_u(t)\quad\forall t.
	\end{equation*}
	
	\subsection{Prosumer load balancing}
\begin{figure}[t!]
		\centering
		\includegraphics[width=0.55\columnwidth]{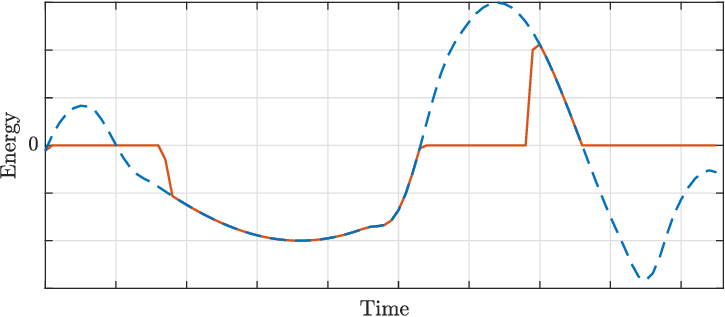}
		\caption{Profile $\rho_u(t)$ (blue dashed) and the resulting profile $\rho_u'(t)$ (red) for a given prosumer $u\in\users_s$.}\label{fig:example_compensation}
	\end{figure}

	In this study, we assume that the prosumers firstly use the storage to balance their own load. In fact, it is convenient to these entities to minimize the electricity bought from the grid. Thus, for a given prosumer $u\in\users_s$, the profile $\rho_u(t)$ is reshaped by using the storage to reduce loads. Such modified profile is denoted by $\rho_u'(t)$. An example of this load balancing is shown in Fig.~\ref{fig:example_compensation}.
	Then, the remaining renewable generation surplus may be further used by the storage system to provide additional flexibility to the community.
	
	Note that this balancing procedure does not affect producers since they do not have loads or consumers since they are not equipped with storage units. So, $\rho_u'(t)=\rho_u(t)$ for all the entities except prosumers in $\users_s$.
	
	Finally, since the prosumer profile is changed, also the maximum energy that can be charged into the storage must be changed accordingly. Specifically, the new energy bound $\maxenergy^{c'}_u(t)$ becomes
	$$
	\maxenergy^{c'}_u(t) = \max\{\rho_u'(t),0\}\quad \forall u\in\users_s, \forall t.
	$$
	
	\subsection{Community level aggregation}
	At a given time $t$, the overall energy demand required by all community entities $L(t)$ is expressed as
	\begin{equation}\label{eq:load}
		L(t) = \sum_{u\in\users}\max\{-\rho_u'(t) ,0\},
	\end{equation}
	whereas the energy generated within the community $R(t)$ is 
	\begin{equation}\label{eq:generation}
		R(t) = \sum_{u\in\users}\max\{\rho_u'(t),0\}.
	\end{equation}
	For each time step $t$, an entity $u$ contributes to $L(t)$ if its load exceeds the generation, i.e., $\rho_u'(t)<0$, while contributes to $R(t)$ if there is a generation surplus, i.e., $\rho_u'(t)>0$. 
	Therefore, the self-consumption at community level $A^0(t)$ is given by
	\begin{equation}\label{eq:self-consumption_0}
		A^0(t) = \min\{L(t),R(t)\},
	\end{equation}
	that represents the energy demand that is matched by the energy generation inside the community. A monetary incentive will be granted to the REC on the basis of this community self-consumption.
	
	Concerning storage units, the stored energy at community level is
	$$
	S(t) = \sum_{u\in\users_s}s_u(t),
	$$
	whereas the related charging and discharging energies are given by
	\begin{align*}
		&E^c(t) = \sum_{u\in\users_s}e^c_u(t),\\
		&E^d(t) = \sum_{u\in\users_s}e^d_u(t).
	\end{align*}
	Let us define
	$$
	\maxEnergy^c(t)=\sum_{u\in\users_s}\maxenergy^{c'}_u(t).
	$$
	Then, at community level, the constraints related to the storage can be rewritten as
	\begin{align}
		& S(t+1) = S(t)+\eta E^c(t) - \frac{1}{\eta}E^d(t)&&\forall t,\label{eq:storage_dyn}\\
		& 0\leq E^c(t)\leq\maxEnergy^c(t)\label{eq:max_c}&&\forall t,\\
		& 0\leq E^d(t)\leq \eta S(t)&&\forall t.\label{eq:max_d}
	\end{align}
	
	Since the storages are charged with energy produced by renewable generators, the net renewable energy injected into the grid at community level (which will be rewarded) is obtained by the following energy balance equation
	\begin{equation}\label{eq:injected_energy}
		G(t) = R(t) - E^c(t) + E^d(t) \quad \forall t.
	\end{equation}
	So, the REC self-consumption in presence of storage at time $t$ is defined as
	\begin{equation}\label{eq:self-consumption_s}
		A^s(t) = \min\{L(t),G(t)\}=\min\{L(t), R(t) - E^c(t) + E^d(t)\}.
	\end{equation}
	
	Hereafter, we will work at community level by employing expressions \eqref{eq:load}-\eqref{eq:self-consumption_s}.
	
	In the next section, it will be shown how to optimally operate the REC storages in order to minimize the REC overall cost.
	
	\section{Optimal Storage Operation}\label{sec:optimization}
	Let $c^p$ and $c^s$ be the energy purchase and selling prices, respectively. Moreover, let $k$ be the unitary incentive price for the self-consumed energy within the community. Suppose to optimize the storage operation in the time interval $\cT = \{0,1,\dots,T-1\}$, the objective function to be minimized is 
	\begin{equation}\label{eq:objective_function}
		\hat{J} = \sum_{\tau\in\cT}\left( c^pL(\tau)-c^s G(\tau) - kA^s(\tau)\right),
	\end{equation} 
	that represents the energy cost of the whole community.
	
	Let us enforce that the storage level at the end of the considered time horizon be equal to that at the initial time. For simplicity, we set such energy level to zero, that is
	\begin{align}
		S(0)=S(T)=0\label{eq:in_fin_cond}.
	\end{align}
	
	Let $\cS$ be the vector containing all the storage charging/discharging control signals in $\cT$, that is
	\begin{align*}
		&\cS = [E^c(0),\dots,E^c(T-1), E^d(0),\dots,E^d(T-1)].
	\end{align*}
	
	Thus, the optimal storage schedule is the minimizer of the following optimization problem.
	\begin{problem}\label{pb:Problem_1}
		\begin{align*}
			\cS^*=\arg&\min_{\cS} \hat{J}\\
			&\mathrm{s.t.}~\eqref{eq:storage_dyn}-\eqref{eq:self-consumption_s},\eqref{eq:in_fin_cond}.
		\end{align*}
	\end{problem}
	
	Let us introduce a proposition which states that $\esc(t)$ and $\esd(t)$ cannot be greater than 0 at the same time.
	
	\begin{proposition}\label{prop:no_both_0}
		Let $\esc(t)$ and $\esd(t)$, $\forall t\in\cT$ be the optimal charging and discharging control signals for Problem~\ref{pb:Problem_1}. Then, 
		\begin{equation}\label{eq:no_both_0}
			\esc(t)\cdot\esd(t)=0,\quad \forall t\in\cT.
		\end{equation}
	\end{proposition}	
	\begin{proof}
		By contradiction assume that at a given time $t$ one has $\esc(t)>0$, $\esd(t)>0$. Suppose $\esd(t)>\esc(t)$. Let us consider the following control signals: $\etd(t)=\esd(t)-\esc(t),~~\etc(t)=0$. By \eqref{eq:injected_energy} one has $G^*(t)=\widetilde{G}(t)$ and then also $A^{s^*}(t)=\widetilde{A}^s(t)$. So, the $t-th$ term of the sum in the objective function \eqref{eq:objective_function} is the same for both solutions. However, the state of charge of the storage at time $t+1$ for the two strategies is
		\begin{align*}
			&S^*(t+1)=S^*(t)+\eta \esc(t)-\frac{1}{\eta}\esd(t),\\
			&\widetilde{S}(t+1)=S^*(t)+\eta \etc(t)-\frac{1}{\eta}\etd(t)= 
			S^*(t)-\frac{1}{\eta}(\esd(t)-\esc(t)).
		\end{align*}
		Then, 
		$$
		\widetilde{S}(t+1)-S^*(t+1)=\frac{1}{\eta}\esc(t)-\eta \esc(t)=\frac{1-\eta^2}{\eta}\esc(t)>0.
		$$
		So, by applying $\etc(t)$ and $\etd(t)$ the cost function till time $t$ is the same as applying $\cS^*$, but the state of charge of the storage is greater. Since \eqref{eq:in_fin_cond} requires that at time $T$ the storage must be empty, such surplus of energy will be sold in the next time steps providing a final cost smaller than that provided by $\cS^*$, leading to a contradiction.\\
		A similar argument holds for the case $\esd(t)<\esc(t)$.
	\end{proof}
	
	\begin{remark}
		It is worthwhile to highlight that, since the community storage is distributed along the community entities, $E^c(t)$ and $E^d(t)$ may be both non-zero for a given feasible solution. However, thanks to Proposition~\ref{prop:no_both_0} the optimal solution is such that the storages are coordinated in such a way that they act as a single aggregated storage of the whole community. In other words, the optimal solution provides a control signal which requires all the storage units of the community to be charged (or discharged) simultaneously, avoiding situations in which a storage is charging when another one is discharging.
	\end{remark}
	
	Since $S(0)=0$, the explicit dynamics of the storage can be written as
	\begin{equation}\label{eq:explicit_S}
		S(t) = \sum_{\tau=0}^{t-1}\left(\eta E^c(\tau) -\frac{1}{\eta}E^d(\tau)\right),\quad \forall t\!\in\!\{0,\dots,T\}.
	\end{equation}
	So, constraint $S(T)=0$ can be represented by 
	\begin{align}
		& \eta^2\sum_{\tau\in\cT} E^c(\tau)=\sum_{\tau\in\cT}E^d(\tau)\label{eq:in_fin_cond_eq}.
	\end{align} 
	By \eqref{eq:injected_energy} and \eqref{eq:in_fin_cond_eq}, the following equation holds
	$$
	\sum_{\tau\in\cT}G(\tau) = \sum_{\tau\in\cT}R(\tau)-\frac{1-\eta^2}{\eta^2}\sum_{\tau\in\cT} E^d(\tau),
	$$
	and hence the objective function \eqref{eq:objective_function} can be written as
	$$
	\hat{J} = \sum_{\tau\in\cT}\left( c^pL(\tau)-c^sR(\tau) + c^s\frac{1-\eta^2}{\eta^2}E^d(\tau) - kA^s(\tau)\right).
	$$
	
	Let us define 
	\begin{equation}\label{eq:alpha}
		\alpha=c^s\frac{1-\eta^2}{\eta^2}>0.
	\end{equation}
	Since neither the energy demand $L(t)$ nor the energy generation $R(t)$ depend on the decision variables, one can aim to minimize the following function
	\begin{equation}\label{eq:objective_function_eq}
		J = \sum_{\tau\in\cT}\left( \alpha E^d(\tau) - k A^s(\tau)\right).
	\end{equation}
	Therefore, the optimal solution of Problem~\ref{pb:Problem_1} coincides with the optimal solution of the following problem.
	\begin{problem}\label{pb:Problem_1_eq}
		\begin{align*}
			\cS^*=\arg&\min_{\cS}~~ J\\
			&\mathrm{s.t.}~\eqref{eq:storage_dyn}-\eqref{eq:self-consumption_s},\eqref{eq:in_fin_cond}.
		\end{align*}
	\end{problem}
	
	To derive the optimal solution of the problem, let focus on the value of $k$, that is on the unitary incentive on the REC self-consumed energy. 
	Depending on the value of $k$, it may be convenient or not to use the storage to optimize the community operation.
	
	The following proposition derives conditions under which the trivial solution $E^d(t)=E^c(t)=0,~\forall t\in\cT$ is optimal.
	
	\begin{proposition}\label{prop:incetive_cond}
		If $k\le \alpha$, then the optimal solution of Problem~\ref{pb:Problem_1_eq} is $E^d(t)=E^c(t)=0,~\forall t\in\cT$.
	\end{proposition}
	\begin{proof}
		Let $E^d(t)=E^c(t)=0,~\forall t\in\cT$ and denote by $J^0$ the value of $J$ for such solution. It holds
		\begin{align*}
			J^0 &= -k \sum_{\tau\in\cT}\! A^s(\tau) = -k\! \sum_{\tau\in\cT}\!\min\{L(\tau),R(\tau)\}= -k\! \sum_{\tau\in\cT}\! A^0(\tau). 
		\end{align*}
		Let us now consider a generic solution. By \eqref{eq:self-consumption_s} and \eqref{eq:self-consumption_0} the following inequality holds
		\begin{align*}
			A^s(t) &=\min\{L(t),R(t)+E^d(t)-E^c(t)\}\\
			&=\min\{L(t)-E^d(t),R(t)-E^c(t)\}+E^d(t)\\
			&\le \min\{L(t),R(t)\}+E^d(t) = A^0(t)+E^d(t).
		\end{align*}
		Thus, the objective function can be bounded from below as follows
		\begin{align*}
			J&\ge  \sum_{\tau\in\cT}\left( \alpha E^d(\tau) - kA^0(\tau) - k E^d(\tau)\right) = \sum_{\tau\in\cT} \left(\alpha-k\right)E^d(\tau) - k \sum_{\tau\in\cT} A^0(\tau). 
		\end{align*}
		Since, by hypothesis $k\le \alpha$, one has
		\begin{align*}
			J\ge -k \sum_{\tau \in \cT} A^0(\tau)=J^0.
		\end{align*}
		Since the solution $E^d(t)=E^c(t)=0,~\forall t\in\cT$ achieves the lower bound, then such a solution is optimal.
	\end{proof}
	
	\begin{remark}
		Proposition~\ref{prop:incetive_cond} provides a threshold value on $k$ denoting the convenience of using the storage. In fact, if $k\le \alpha$ the best solution does not require the usage of the storage to minimize the REC cost. For this reason, from now on we will focus on the case $k>\alpha$. 
	\end{remark}

	Let us introduce two lemmas which will be instrumental in proving the main theorems.
	
	\begin{lemma}\label{lem:main_in_avanti}
		Let $\esc(t)$ and $\esd(t)$, $\forall t\in\cT$ be the optimal charging and discharging control signals for Problem~\ref{pb:Problem_1_eq}. Then, choose $t_1\in\cT$ such that $\esc(t_1)>0$ and set $t_2=\min\{t>t_1\colon \esd(t)>0\}$.  Suppose $0<\eps<\min\{\esc(t_1),\frac{1}{\eta^2}\esd(t_2)\}$. Let $\widetilde\cS$ be a solution composed by the following charging and discharging control signals:
		\begin{align*}
			\etc(t_1)&=\esc(t_1)-\eps,       & \etc(t)&=\esc(t),~\forall t\neq t_1,\\
			\etd(t_2)&=\esd(t_2)-\eta^2 \eps,& \etd(t)&=\esd(t),~\forall t\neq t_2,
		\end{align*}
		Then, $\widetilde\cS$ is a feasible solution for Problem~\ref{pb:Problem_1_eq}.
	\end{lemma}	
	\begin{proof}
		First, notice that $t_2$ always exists. In fact, since $S^*(t_1+1)>0$ and $S^*(T)=0$, there exists a time $t=t_1+1,\ldots,T-1$ where the storage is discharged. 
		
		Now, let us prove the feasibility of $\widetilde\cS$. For any time $t<t_1$ the two solutions are identical. At time $t_1$ one has $0<\etc(t_1)<\esc(t_1)$ and by Proposition~\ref{prop:no_both_0} $\etd(t_1)=\esd(t_1)=0$. So, the charging control signal at $t_1$ is feasible. For any time $t_1<t<t_2$ both solutions involve the same charging control signals, while $\etd(t)=\esd(t)=0$. 
		At time $t_2$, the storage is discharged by $\etd(t_2)$. In order to be feasible, we must guarantee $\etd(t_2)\le\eta \widetilde S(t_2)$, according to \eqref{eq:max_d}.
		Since $\etd(t)=0$ for $t=t_1+1,\ldots,t_2-1$, one has 
		\begin{align*}
			\widetilde S(t_2)&=S^*(t_1)+\eta \etc(t_1)+\eta\sum_{\tau=t_1+1}^{t_2-1} \esc(\tau)=S^*(t_1)+\eta\sum_{\tau=t_1}^{t_2-1} \esc(\tau)-\eta\eps.
		\end{align*}
		So,
		\begin{align*}
			&\widetilde S(t_2+1)~=~S^*(t_1)+\eta\sum_{\tau=t_1}^{t_2-1} \esc(\tau)-\eta\eps-\frac{1}{\eta}\etd(t_2)\\
			&~~~=S^*(t_1)+\eta\sum_{\tau=t_1}^{t_2-1} \esc(\tau)-\eta\eps-\frac{1}{\eta}(\esd(t_2)-\eta^2\eps)\\
			&~~~=S^*(t_1)+\eta\sum_{\tau=t_1}^{t_2-1} \esc(\tau)-\frac{1}{\eta}\esd(t_2)=S^*(t_2+1).
		\end{align*}
		Since $\widetilde S(t_2+1)=S^*(t_2+1)$, and because the two solutions are identical from time $t_2+1$ onwards, $\widetilde\cS$ is a feasible solution for Problem~\ref{pb:Problem_1_eq}.
	\end{proof}	
	
	\begin{lemma}\label{lem:main_all_indietro}
		Let $\esc(t)$ and $\esd(t)$, $\forall t\in\cT$ be the optimal charging and discharging control signals for Problem~\ref{pb:Problem_1_eq}. Then, choose $t_2\in\cT$ such that $\esd(t_2)>0$ and set $t_1=\max\{t<t_2\colon \esc(t)>0\}$. Suppose $0<\eps<\min\{\eta^2 \esc(t_1),\esd(t_2)\}$. Let $\widetilde\cS$ be a solution composed by the following charging and discharging control signals:
		\begin{align*}
			\etc(t_1)&=\esc(t_1)-\frac{1}{\eta^2}\eps,       & \etc(t)&=\esc(t),~\forall t\neq t_1,\\
			\etd(t_2)&=\esd(t_2)-\eps,& \etd(t)&=\esd(t),~\forall t\neq t_2,
		\end{align*}
		Then, $\widetilde\cS$ is a feasible solution for Problem~\ref{pb:Problem_1_eq}.
	\end{lemma}	
	\begin{proof}
		The proof follows the same reasoning as that of Lemma~\ref{lem:main_in_avanti}.
	\end{proof}	
	
	\begin{theorem}\label{th:main_1}
		Let us consider Problem~\ref{pb:Problem_1_eq}, and let $\esc(t)$ and $E^{d^*}(t)$ be the optimal charging and discharging control signals for $\forall t \in \cT$. It holds that
		\begin{align}
			\esc(t)=0 & \text{~~~~if } L(t)\ge R(t),\label{eq:main_L_bigger}\\
			\esd(t)=0 & \text{~~~~if } L(t)\le R(t).\label{eq:main_R_bigger}
		\end{align}
	\end{theorem}
	\begin{proof}
		First, let us prove \eqref{eq:main_L_bigger}. Let $t_1\in\cT$ be such that $L(t_1)\ge R(t_1)$. By contradiction assume $\esc(t_1)>0$. Let $\widetilde\cS$ be defined as in Lemma~\ref{lem:main_in_avanti}, so it is a feasible solution for Problem~\ref{pb:Problem_1_eq}.
		
		Let $J^*$ be the optimal cost of Problem~\ref{pb:Problem_1_eq}, while $\widetilde{J}$ be the cost related to $\widetilde\cS$. Notice that $J^*$ and $\widetilde J$ differ only in the terms depending on $t_1$ and $t_2$. Moreover, by Proposition~\ref{prop:no_both_0}, $\esd(t_1)=\esc(t_2)=\etd(t_1)=\etc(t_2)=0$. Thus,
		\begin{align*}
			&J^*-\widetilde J=-k\left(A^{s^*}(t_1)-\widetilde{A}^s(t_1)\right) +\alpha\left(\esd(t_2) -\etd(t_2) \right) -k\left(A^{s^*}(t_2)-\widetilde{A}^s(t_2)\right).
		\end{align*}
		Since $L(t_1)\ge R(t_1)$, by \eqref{eq:self-consumption_s} one has
		\begin{align*}
			&A^{s^*}(t_1)=R(t_1)-\esc(t_1),\\& \widetilde{A}^{s}(t_1)=R(t_1)-\etc(t_1)=R(t_1)-\esc(t_1)+\eps.
		\end{align*}
		and hence
		$$
		J^*-\widetilde J=k\eps+\alpha\eta^2\eps-k\left(A^{s^*}(t_2)-\widetilde{A}^s(t_2)\right).
		$$
		
		By definition
		\begin{align*}
			\widetilde{A}^s(t_2)&=\min\{L(t_2),R(t_2)+\etd(t_2)\}\\
			&=\min\{L(t_2),R(t_2)+\esd(t_2)-\eta^2\eps\}\\
			&=\min\{L(t_2)+\eta^2\eps,R(t_2)+\esd(t_2)\}-\eta^2\eps\\&\ge A^{s^*}(t_2)-\eta^2\eps.
		\end{align*}
		Then,
		$$
		J^*-\widetilde J\ge k\eps+\alpha\eta^2\eps-k \eta^2\eps=k\eps(1-\eta^2)+\alpha\eta^2\eps>0.
		$$
		Since $J^*>\widetilde J$ a contradiction occurs.
		
		To prove \eqref{eq:main_R_bigger}, a specular reasoning can be repeated by exploiting Lemma~\ref{lem:main_all_indietro}.
	\end{proof}
	Clearly, Theorem~\ref{th:main_1} states that a candidate solution to Problem~\ref{pb:Problem_1_eq} must satisfy \eqref{eq:main_L_bigger}-\eqref{eq:main_R_bigger} that automatically enforces condition \eqref{eq:no_both_0}.
	
	\begin{theorem}\label{th:main_2}
		Let us consider Problem~\ref{pb:Problem_1_eq}, and let $\esc(t)$ and $E^{d^*}(t)$ be the optimal charging and discharging control signals for $\forall t \in\cT$. It holds that
		\begin{align}
			&\esc(t)\le R(t)-L(t)~~\text{ if } L(t)\le R(t),\label{eq:main_2_R_bigger}\\
			&\esd(t)\le L(t)-R(t)~~\text{ if } L(t)\ge R(t).\label{eq:main_2_L_bigger}
		\end{align}
	\end{theorem}
	\begin{proof}
		First, let us prove \eqref{eq:main_2_R_bigger}. Let $t_1\in\cT$ be such that $ L(t_1)\le R(t_1)$. By contradiction assume $\esc(t_1)>R(t_1)-L(t_1)$. Let $\widetilde\cS$ be defined as in Lemma~\ref{lem:main_in_avanti} and assume $0<\eps<\min\{\esc(t_1),\frac{1}{\eta^2}\esd(t_2),\esc(t_1)-R(t_1)+L(t_1)\}$. Since $R(t_1)\ge L(t_1)$ one has $\esc(t_1)>\esc(t_1)-R(t_1)+L(t_1)$. So, choosing $\eps$ such that $0<\eps<\min\{\frac{1}{\eta^2}\esd(t_2),\esc(t_1)-R(t_1)+L(t_1)\}$ leads to a feasible solution to Problem~\ref{pb:Problem_1_eq}.
		
		Let $J^*$ be the optimal cost of Problem~\ref{pb:Problem_1_eq}, while $\widetilde{J}$ be the cost related to $\widetilde\cS$. Notice that $J^*$ and $\widetilde J$ differ only in the terms depending on $t_1$ and $t_2$. Moreover, by Proposition~\ref{prop:no_both_0}, $\esd(t_1)=\esc(t_2)=\etd(t_1)=\etc(t_2)=0$. Thus,
		\begin{align*}
			&J^*-\widetilde J=-k\left(A^{s^*}(t_1)-\widetilde{A}^s(t_1)\right) +\alpha\left(\esd(t_2)-\etd(t_2) \right)-k\left(A^{s^*}(t_2)-\widetilde{A}^s(t_2)\right).
		\end{align*}
		Since $\esc(t_1)>R(t_1)-L(t_1)$ one has
		\begin{align*}
			A^{s^*}(t_1) &= \min\{L(t_1),R(t_1)-\esc(t_1)\}\\
			& =\min\{L(t_1)-R(t_1),-\esc(t_1)\}+R(t_1) \\&= R(t_1)-\esc(t_1).
		\end{align*} 
		Recalling that $\eps < \esc(t_1)-R(t_1)+L(t_1) $
		\begin{align*}
			\widetilde{A}^s(t_1) &= \min\{L(t_1),R(t_1)-\etc\!(t_1)\}\\
			& =\min\{L(t_1)-R(t_1),-\esc\!(t_1)+\eps\}\!+\!R(t_1)\\
			& = \min\{\esc\!(t_1)-R(t_1)+L(t_1),\eps\}\!+\!R(t_1)\!-\!\esc\!(t_1)\\
			& = R(t_1)-\esc\!(t_1)+\eps.
		\end{align*} 
		Thus $	A^{s^*}(t_1) - \widetilde{A}^s(t_1) = -\eps$ and hence
		$$ J^*-\widetilde J=k\eps+\alpha\eta^2\eps-k\left(A^{s^*}(t_2)-\widetilde{A}^s(t_2)\right).$$
		By definition
		\begin{align*}
			\widetilde{A}^s(t_2)&=\min\{L(t_2),R(t_2)+\etd(t_2)\}\\
			&=\min\{L(t_2),R(t_2)+\esd(t_2)-\eta^2\eps\}\\
			&=\min\{L(t_2)+\eta^2\eps,R(t_2)+\esd(t_2)\}-\eta^2\eps\\&\ge A^{s^*}(t_2)-\eta^2\eps.
		\end{align*}
		Then,
		$$
		J^*-\widetilde J\ge k\eps+\alpha\eta^2\eps-k \eta^2\eps=k\eps(1-\eta^2)+\alpha\eta^2\eps>0.
		$$
		Since $J^*>\widetilde J$ a contradiction occurs.
		To prove \eqref{eq:main_2_L_bigger}, a specular reasoning can be repeated by exploiting Lemma~\ref{lem:main_all_indietro} and by putting $\eps<\esd(t_2)-L(t_2)+R(t_2)$.
	\end{proof}
	
	\begin{corollary}\label{cor:A_s}
		Let $\cS^*$ be the optimal solution of Problem~\ref{pb:Problem_1_eq}. Then,	
		\begin{equation}\label{eq:opt_A}
			\begin{aligned}
				A^{s^*}(t) &=A^0(t)+\esd(t)\\&=\begin{cases}
					L(t)\!\!\! & \text{ if }L(t)\le R(t)\\
					R(t)+\esd(t)\!\!\! & \text{ if }L(t)\ge R(t).
				\end{cases}
			\end{aligned}
		\end{equation}
	\end{corollary}
	\begin{proof}
		Consider the case $L(t)\le R(t)$. By \eqref{eq:main_R_bigger}, $\esd(t)=0$, and by Theorem~\ref{th:main_2}, one has $\esc(t)\le R(t)-L(t)$. Thus,
		$$
		A^{s^*}(t)=\min\{L(t),R(t)-\esc(t)\}=L(t).
		$$
		Let us analyze the case $L(t)\ge R(t)$. By \eqref{eq:main_L_bigger}, $\esc(t)=0$ and hence 
		$$
		A^{s^*}(t)=\min\{L(t),R(t)+\esd(t)\}.
		$$
		By Theorem~\ref{th:main_2}, $\esd(t)\le L(t)-R(t)$ and hence $A^{s^*}(t)=R(t)+\esd(t)$.
	\end{proof}
	
	Let us partition the time indices in two sets, depending on the fact that the REC load is less or greater than generation:
	\begin{align*}
		\Lm&=\{t\in\cT\colon L(t)\le R(t)\},\\
		\Lp&=\{t\in\cT\colon L(t)> R(t)\}.
	\end{align*}
	
	By Corollary~\ref{cor:A_s}, and by adding \eqref{eq:main_L_bigger}-\eqref{eq:main_2_L_bigger} to the constraints of Problem~\ref{pb:Problem_1_eq}, the objective function in \eqref{eq:objective_function_eq} can be rewritten as
	\begin{align*}
		J&=\alpha\sum_{\tau\in\cT} E^d(\tau)-k\sum_{\tau\in\cT}\left( A^0(\tau)+ E^d(\tau)\right)=-k\sum_{\tau\in\cT} A^0(\tau)+ (\alpha-k)\sum_{\tau\in\Lp} E^d(\tau),
	\end{align*}
	where the last equality comes from \eqref{eq:main_R_bigger}.
	
	Since $A^0(t)$ is known and constant, and $k>\alpha$, the optimal solution of Problem~\ref{pb:Problem_1_eq} coincides with the optimal solution of the following problem.
	\begin{problem}\label{pb:Problem_max}
		\begin{align}
			\cS^*&=\arg\max_{\cS} \sum_{t\in\Lp} E^d(t)\\
			&~~\mathrm{s.t.}\nonumber\\
			& \begin{aligned}
				S(t+1) = S(t)+\eta E^c(t) - \frac{1}{\eta}E^d(t)
			\end{aligned}&\forall t\in\cT,\\
			& 0\leq E^c(t)\leq\maxEnergy^c(t)&\forall t\in\cT,\label{eq:E_c_constr}\\
			& 0\leq E^d(t)\leq \eta S(t)&\forall t\in\cT,\label{eq:E_d_constr}\\
			& E^c(t) \le R(t)-L(t),~~  E^d(t) = 0 & \forall t\in\Lm,\label{eq:max_E_c_fin}\\
			& E^d(t) \le L(t)-R(t),~~ E^c(t) = 0 & \forall t\in\Lp,\label{eq:max_E_d_fin}\\
			& S(0)=S(T)=0.
		\end{align}
	\end{problem}
	Note that the optimality conditions reported in Theorems~\ref{th:main_1} and \ref{th:main_2} are summarized by constraints \eqref{eq:max_E_c_fin}-\eqref{eq:max_E_d_fin}.
	
	Since the objective function of Problem~\ref{pb:Problem_max} involves only $E^d(t)$, it is apparent that the optimal solution maximizes the overall discharged energy. Operatively, at a given time $t\in\Lp$ the optimal control signal is to choose $E^d(t)$ as bigger as possible. Thanks to \eqref{eq:E_d_constr}-\eqref{eq:max_E_d_fin}, the optimal solution results
	\begin{equation}\label{eq:opt_E_d}
		\esd(t)=\min\{L(t)-R(t),\eta S(t)\},~~\esc(t)=0,\quad \forall t\in\Lp.
	\end{equation}
	On the contrary, when $t\in\Lm$ by \eqref{eq:main_R_bigger} one has $E^d(t)=0$. Note that the choice of the control variables at these time steps does not influence the objective function. Thus, the optimal charging control signal will be that maximizing the stored energy while satisfying the constraints.
	
	By \eqref{eq:in_fin_cond_eq} it holds
	$$
	\sum_{\tau=0}^{t-1}E^c(\tau)+ E^c(t)+\sum_{\tau=t+1}^{T-1}E^c(\tau)=
	\sum_{\tau\in\cT}\left(\frac{E^d(\tau)}{\eta^2}\right),\\
	$$
	and hence
	\begin{align*}
		E^c(t)&=\sum_{\tau=0}^{t-1}\left(\frac{E^d(\tau)}{\eta^2}-E^c(\tau)\right)\\&+                 \sum_{\tau\in\cT}\left(\frac{E^d(\tau)}{\eta^2}\right) -
		\sum_{\tau=t+1}^{T-1}E^c(\tau).	
	\end{align*}
	By \eqref{eq:explicit_S}, one has
	$$
	-\frac{S(t)}{\eta}=\sum_{\tau=0}^{t-1}\left(\frac{E^d(\tau)}{\eta^2}-E^c(\tau)\right).
	$$
	Since $E^c(t)\ge 0$, $\forall t$, it follows
	$$
	E^c(t)\le-\frac{S(t)}{\eta}+                 \sum_{\tau=t}^{T-1}\left(\frac{E^d(\tau)}{\eta^2}\right).
	$$
	Moreover, by \eqref{eq:E_d_constr} and \eqref{eq:max_E_d_fin} one has $0\le E^d(t)\le L(t)-R(t)$, and then
	\begin{equation}\label{eq:ub_c}
		E^c(t)\le-\frac{S(t)}{\eta}+\sum_{\tau\in\Lp,\tau>t}\frac{L(\tau)-R(\tau)}{\eta^2}.
	\end{equation}
	So, by \eqref{eq:E_c_constr}, \eqref{eq:max_E_c_fin} and \eqref{eq:ub_c} the maximum, and thus the optimal, charging control signal is
	\begin{equation}\label{eq:opt_E_c}
		\esc\!\!(t)=\min\Bigg\{\maxEnergy^c\!(t),\mathrm{P}(t),-\frac{S(t)}{\eta}-\!\!\!\sum_{\tau\in\Lp,\tau>t}\!\!\frac{\mathrm{P}(\tau)}{\eta^2}\Bigg\},
	\end{equation}
	where $\mathrm{P}(t) = R(t)-L(t)$.
	
	\section{Numerical Results}\label{sec:results}
	To show how the proposed solution performs according to a given setup, two illustrative examples involving small and large renewable generation are provided. 
	The optimization period spans over 24 hours with a sampling time 15 minutes, i.e., $T=96$.
	A simplified community structure involving one consumer, one prosumer and one producer is considered. 
	Both the prosumer and the producer are assumed to be equipped with a storage unit whose efficiency is set to 0.9.
	Grid prices are supposed to be $c^p = 0.35$~\euro/kWh and $c^s = 0.18$~\euro/kWh, whereas the incentive $k=0.12$~\euro/kWh.
	Note that, since $\alpha = c^s(1-\eta^2)/\eta^2=0.042$~\euro/kWh, from Proposition~\ref{prop:incetive_cond} the incentive $k$ is chosen to foster storage utilization. 
	
	\subsection{Small renewable generation}
\begin{figure}[t]
		\centering
		\includegraphics[width=0.5\columnwidth]{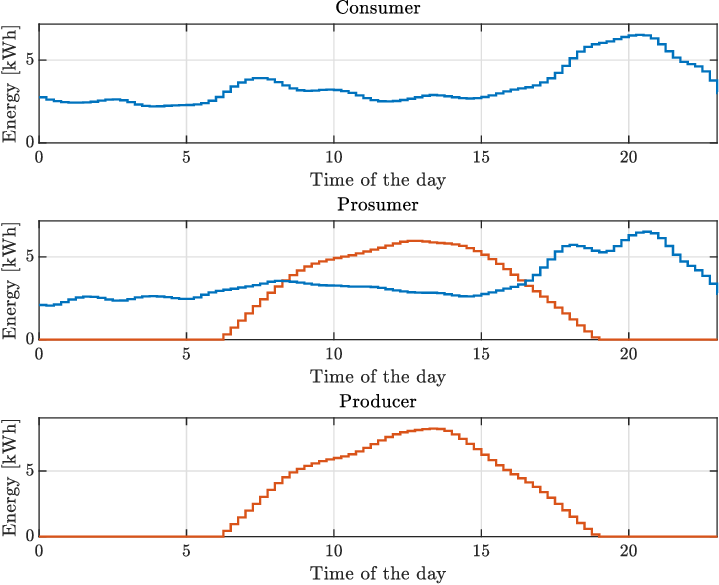}
		\caption{Load (blue) and generation (red) profiles of the community entities.}\label{fig:entity_profile_small}
	\end{figure}

\begin{figure}[t]
		\centering
		\includegraphics[width=0.5\columnwidth]{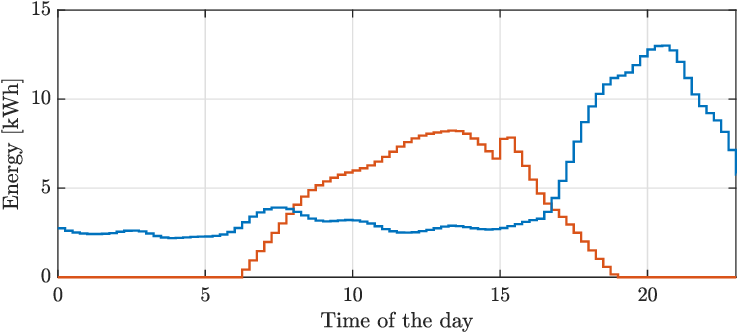}
		\caption{Load (blue) and generation (red) profiles at community level in case of low renewable generation.}\label{fig:community_profile}
	\end{figure}

	In this example, a community involving a daily renewable generation lower than the daily load is considered. Profiles of load and generation of the community entities are depicted in Fig.~\ref{fig:entity_profile_small}.
	Aggregating at community level, community load and generation are computed and reported in Fig.~\ref{fig:community_profile}.
	The aggregated profiles show a surplus generation during the central hours of the day, and an excess of load during the other periods. Note that the generation peak occurring around 15:00 is due to the load balancing performed by the prosumer. 
	
\begin{figure}[t]
		\centering
		\includegraphics[width=0.5\columnwidth]{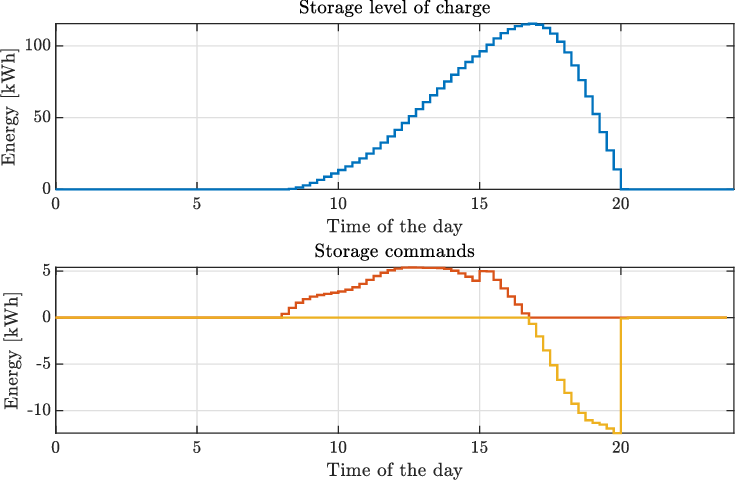}
		\caption{Storage level of charge (blue) and related charging (red) and discharging (yellow) control signals in case of low renewable generation.}\label{fig:storage_small}
	\end{figure}

\begin{figure}[t]
		\centering
		\includegraphics[width=0.5\columnwidth]{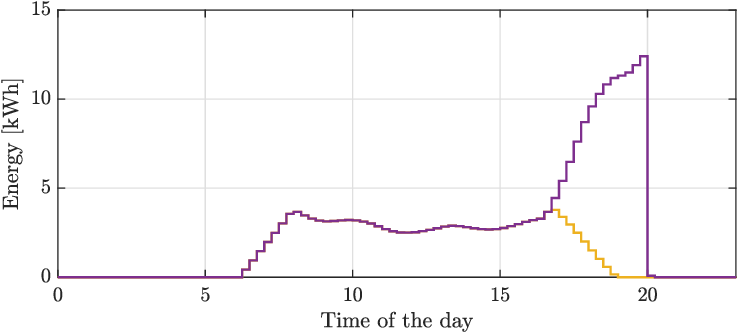}
		\caption{Self-consumption profiles $A^0(t)$ (yellow) and $A^s(t)$ (purple) in case of low renewable generation.}\label{fig:self_consumption_small}
	\end{figure}

	By exploiting \eqref{eq:opt_E_d} and \eqref{eq:opt_E_c}, the energy surplus is fully employed to compensate part of the load in the last period of the day. This behavior is evident by looking at the storage dynamics and the related charging and discharging control signals in Fig.~\ref{fig:storage_small}.
	Moreover, as stated by Corollary~\ref{cor:A_s}, the obtained storage schedule is such that the self-consumption $A^0(t)$ is maintained/enhanced when the storage is charging/discharging, as shown in Fig.~\ref{fig:self_consumption_small}.
	Thus the obtained profile follows the load until the storage is fully discharged. The storage operation in this setup is capable of reducing the community cost of about 8\%, while the incentive gained is increased by about 1.77 times.
	Community cost and incentive concerning both setups are summarized in Tab.~\ref{tab:cost}. 
	\begin{table}[b]
		\centering
		\begin{tabular}{c|c|c|c|c|}
			\cline{2-5}
			& \multicolumn{2}{|c|}{Low generation} & \multicolumn{2}{|c|}{High generation}\\	\cline{2-5}
			& No storage & Optimal & No storage & Optimal \\ \hline
			\multicolumn{1}{|l|}{Cost [\euro]} &  96.61 & 88.53 &  47.99
			& 29.61\\ \hline
			\multicolumn{1}{|l|}{Incentive [\euro]} & 16.24 & 28.70 & 18.73 & 47.10\\ \hline
		\end{tabular}\vspace{6pt}\caption{Community cost and incentive}\label{tab:cost}
	\end{table}
	
	\subsection{Large renewable generation}
	\begin{figure}[t]
		\centering
		\includegraphics[width=0.5\columnwidth]{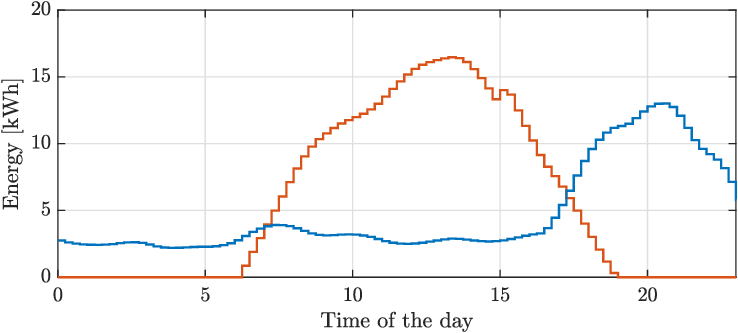}
		\caption{Load (blue) and generation (red) profiles at community level in case of high renewable generation.}\label{fig:community_profile_big}
	\end{figure}
	\begin{figure}[t]
		\centering
		\includegraphics[width=0.5\columnwidth]{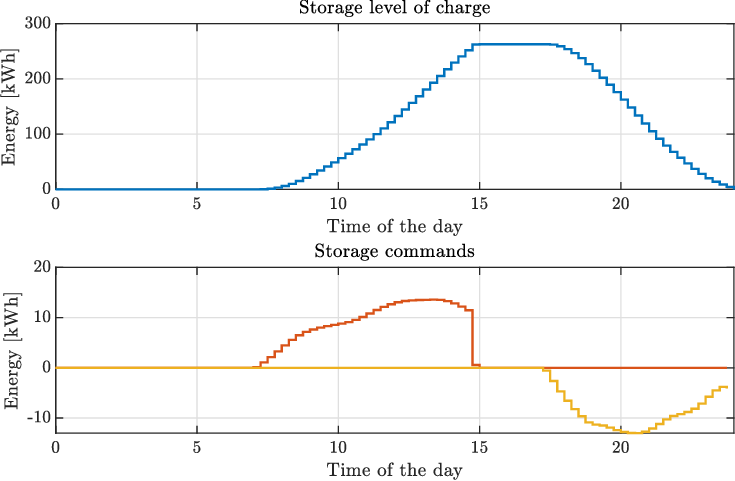}
		\caption{Storage dynamics (blue) and related charging (red) and discharging (yellow) control signals in case of high renewable generation.}\label{fig:storage_big}
	\end{figure}
	\begin{figure}[t]
		\centering
		\includegraphics[width=0.5\columnwidth]{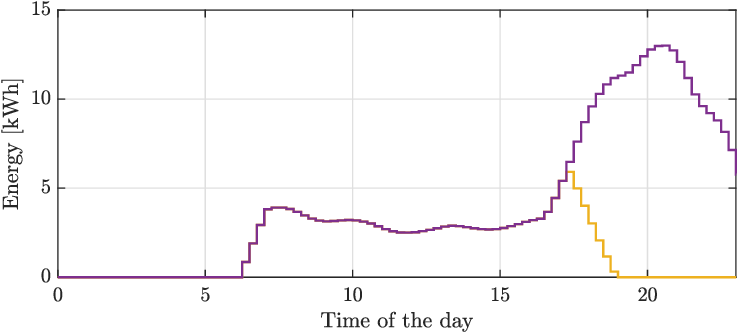}
		\caption{Self-consumption profiles $A^0(t)$ (yellow) and $A^s(t)$ (purple) in case of high renewable generation.}\label{fig:self_consumption_big}
	\end{figure}
	In this scenario, we consider the same load and generation profiles as before except for the producer, whose production is doubled. The aggregated profiles are reported in Fig.~\ref{fig:community_profile_big}.
	Clearly, as reported in Tab.~\ref{tab:cost}, the community cost is much lower than the previous case even without the presence of the storage units. However, the optimal storage management leads to a cost that is further reduced by about 38\%, while the incentive is increased by about 2.5 times. The battery is managed so that the energy surplus is used to meet the load excess in the last period of the day. The storage solution and the self-consumption profiles are reported in Figs.~\ref{fig:storage_big} and~\ref{fig:self_consumption_big}.
	Even though the generation is much higher than the previous case, the storage is charged by only the energy necessary to fulfill the load at the end of the day. The rest of the generated energy is sold to the grid.
	
	\section{Conclusions}\label{sec:conclusion}
	In this work, an optimization procedure for REC storage operation is proposed. Firstly, the load/generation profiles of the community entities are merged on the basis of member typology. Then, the problem is formulated as a linear program where the objective is to minimize the community cost by exploiting the storage units installed within the community entities. To derive an optimal solution, necessary conditions for the optimal storage scheduling are devised. Specifically, the optimal storage operation is that which maximizes the community self-consumption. Numerical results show a consistent cost reduction with respect to scenarios not involving storage units. Most notably, in both the considered examples, the community self-consumption is considerably increased when the proposed method is involved, leading to several environmental benefits. Future research directions will be focused on handling the uncertainty affecting load and generation, as well as on adapting the proposed framework in more complex scenarios that may involve electric vehicles, shiftable loads, and/or demand response programs.

\bibliography{community}
\bibliographystyle{ieeetran}
\end{document}